\documentclass[conference]{IEEEtran}
\IEEEoverridecommandlockouts
\usepackage{cite}
\usepackage{amsmath,amssymb,amsfonts}
\usepackage{algorithmic}
\usepackage{graphicx}
\usepackage{textcomp}
\usepackage{xcolor}

\usepackage{braket} 
\usepackage[breaklinks]{hyperref} 
\usepackage{authblk}

\def\BibTeX{{\rm B\kern-.05em{\sc i\kern-.025em b}\kern-.08em
    T\kern-.1667em\lower.7ex\hbox{E}\kern-.125emX}}
    
\usepackage{booktabs} 
\usepackage{multirow}
\usepackage{booktabs}
\usepackage{graphicx}  
\usepackage{diagbox}
\usepackage[braket]{qcircuit} 
\usepackage{threeparttable} 
\usepackage{stfloats} 
\usepackage{enumitem} 
\usepackage{fancyheadings} 
\usepackage{wrapfig} 

\usepackage{times}
\usepackage{graphicx}
\usepackage{epsf}
\usepackage{verbatim}
\usepackage{url}
\usepackage{color}
\usepackage{alltt}

\newcommand{\Caption}{\caption}

\newcommand{\CodeIn}[1]{{\small\texttt{#1}}}

\newcommand{\Comment}[1]{}

\newenvironment{CodeOut}{\begin{scriptsize}}{\end{scriptsize}}

\newcommand{\SmallSpace}{\vspace*{-1.4ex}}

\author{Junjie Luo*\thanks{*This work was carried out when Junjie Luo was at Kyushu University.}}
\author{Pengzhan Zhao}
\author{Zhongtao Miao}
\author{Shuhan Lan}
\author{Jianjun Zhao}
\affil{Kyushu University, Japan}

\begin{document}

\title{A Comprehensive Study of Bug Fixes in Quantum Programs  
}

\maketitle

\begin{abstract}
As quantum programming evolves, more and more quantum programming languages are being developed. As a result, debugging and testing quantum programs have become increasingly important. While bug fixing in classical programs has come a long way, there is a lack of research in quantum programs. To this end, this paper presents a comprehensive study on bug fixing in quantum programs. We collect and investigate 96 real-world bugs and their fixes from four popular quantum programming languages (\texttt{Qiskit}, \texttt{Cirq}, \texttt{Q\#}, and \texttt{ProjectQ}). 
Our study shows that a high proportion of bugs in quantum programs are quantum-specific bugs (over 80\%), which requires further research in the bug fixing domain. We also summarize and extend the bug patterns in quantum programs and subdivide the most critical part, math-related bugs, to make it more applicable to the study of quantum programs. Our findings summarize the characteristics of bugs in quantum programs and provide a basis for studying testing and debugging quantum programs. 
\end{abstract}


\section{Introduction}
Debugging and testing are critical parts of an integrated software development method in modern software development. A software bug is regarded as the abnormal program behaviors which deviate from its specification~\cite{allen2002bug}, including poor performance when a threshold level of performance is included as part of the specification. Software defects significantly impact the economy, security, and quality of life, yet relying on manual diagnosis and repair of software bugs consumes significant time and money. To address this problem, many software engineering tasks, such as program analysis, debugging, and software testing, are devoted to developing techniques and tools for locating and fixing bugs. In addition, software bugs can be handled or prevented more effectively by studying past bugs and their fixes.

Quantum programming is the process of designing and constructing executable quantum programs to achieve a specific computational result. Recently, many quantum programming languages have been developed for the development of quantum software, for instance, \texttt{Qiskit}~\cite{ibm2021qiskit}, \texttt{Cirq}~\cite{cirq2018google}, \texttt{Q\#}~\cite{svore2018q}, and \texttt{ProjectQ}~\cite{projectq2017projectq}. The current research in quantum programming focuses mainly on problem analysis, language design, and implementation. However, despite their importance, program debugging and software testing have received relatively little attention in the quantum programming paradigm. The specific features introduced in quantum programming, such as superposition, entanglement, and no-cloning, make them different for bug fixing compared to classical programs. Although several approaches have been proposed for debugging and testing quantum software~\cite{li2020projection,huang2019statistical,honarvar2020property,miranskyy2019testing,ali2021assessing} recently, no research work has been focused on bug fixing in quantum programs~\cite{zhao2020quantum}.

This paper first proposes five questions that need analysis and discussion. On this basis, we collect and investigate 96 real-world bugs and their fixes from four popular quantum programming languages (\texttt{Qiskit}, \texttt{Cirq}, \texttt{Q\#}, and \texttt{ProjectQ}). Finally, we obtained nine findings to answer the questions. Our main findings can be summarized as follows:

\begin{itemize}[leftmargin=2em]
\setlength{\itemsep}{3pt}
  \item More than 80\% of the bugs we collected is quantum-specific, which means that classical bug fixing methods are insufficient, and bugs in quantum programs need further research. 
  \item Although the current bugs in quantum programs are generally simple, the high complexity bugs are all quantum-specific bugs, which means bug fixing for quantum programs is challenging. 
  \item We summarize and expand the bug patterns in quantum programs and count the number of bugs for each bug pattern. Math-related bug patterns are the focus of our analysis, which are uncommon in classical programs. 
\end{itemize}

We believe that the results of our study can benefit future research in two aspects:

\begin{enumerate}
\setlength{\itemsep}{3pt}
  \item It can provide insights into how to develop and improve the bug fixing methods for existing quantum programs.
  For example, what knowledge is needed to repair quantum programs and whether this knowledge is included in classical program fixing methods. In addition, the results also show the potential regarding the application of existing automatic fixing methods in quantum programs.
  \item It can provide a more precise direction for studying quantum programs bugs. For example, the results show some bug patterns in quantum programs and analyze the frequency of each bug pattern. Based on these bug patterns that we have summarized, our future work can propose new methods for locating quantum-specific bugs, as well as developing static analysis tools. 
\end{enumerate}

The rest of the paper is organized as follows. Section~\ref{sec:background} briefly introduces the background knowledge of quantum programming. Section~\ref{sec:methodology} describes the research methodology of our comprehensive study. Section~\ref{sec:result} discusses some findings of our study. Related work is discussed in Section~\ref{sec:related-work}, and concluding remarks are given in Section~\ref{sec:conclusion}.

\section{Background}
\label{sec:background}

This section briefly introduces some basic concepts on quantum computing
as well as quantum programming frameworks. 

\subsection{Basic Concepts}

A quantum bit (qubit) is the analog of one classical bit but has many different properties. A classical bit, like a coin, has only two states, 0 and 1, while a qubit can be in a continuum of states between $\Ket{0}$ and $\Ket{1}$ in which the $\Ket{}$ notation is called Dirac notation. We can represent a qubit mathematically as $\Ket{\psi} = \alpha\Ket{0} + \beta\Ket{1}$ where $|\alpha|^2 + |\beta|^2 = 1$ and the numbers $\alpha$ and $\beta$ are complex numbers. The states $\Ket{0}$ and $\Ket{1}$ are called computational basis states. Unlike classical bits, we cannot examine a qubit directly to get the values of $\alpha$ and $\beta$. Instead, we measure a qubit to obtain either the result 0 with probability $|\alpha|^2$ or the result 1 with probability $|\beta|^2$.

Quantum gates are used to do the quantum computation, which means manipulating quantum information. Some basic quantum gates are as follows:
\begin{itemize}[leftmargin=2em]
\setlength{\itemsep}{3pt}
  \item Quantum NOT gate takes the state $\Ket{\psi} = \alpha\Ket{0} + \beta\Ket{1}$ into the state $\Ket{\psi} = \alpha\Ket{1} + \beta\Ket{0}$. We can use a matrix to represent this operation:
  \begin{gather*}X = \begin{bmatrix} 0 & 1\\1 & 0\end{bmatrix}\end{gather*}
  \item The Z gate can be expressed as 
  \begin{gather*}Z = \begin{bmatrix} 1 & 0 \\ 0 & -1\end{bmatrix}\end{gather*} 
  From the matrix, we know the Z gate leaves the $\Ket{0}$ unchanged and changes the sign of $\Ket{1}$.
  \item The Hadamard gate turns the $\Ket{0}$ into $(\Ket{0} + \Ket{1})/\sqrt{2}$ and turns the $\Ket{1}$ into $(\Ket{0} - \Ket{1})/\sqrt{2}$. The matrix form of the Hadamard gate is 
  \begin{gather*}H = \frac{1}{\sqrt{2}}\begin{bmatrix} 1 & 1\\ 1 & -1 \end{bmatrix}\end{gather*}
\end{itemize}

All the matrices are unitary ones. Besides these single-qubit gates, there are multiple qubit gates, such as the Controlled-NOT gate (CNOT gate). This gate has two input qubits, the control qubit and the target qubit. If the control qubit is 0, then the target qubit remains unchanged. If the control qubit is 1, then the target qubit is flipped. We can express the behavior of the CNOT gate as $\Ket{A, B} \rightarrow \Ket{A, B \oplus A}$. 

Quantum circuits are models of all kinds of quantum processes. We can build quantum circuits with quantum gates and use wires to connect the components in quantum circuits. These wires can represent the passage of time or a physical particle moving from one position to another. Another essential operation in quantum circuits is measurement. Measurement operation observes a single qubit and obtains a classic bit with a certain probability. 
Nielsen's book\cite{nielsen2002quantum} has a more detailed explanation of quantum computation.

\subsection{Quantum Programming Frameworks}

Recently, several open-source programming frameworks for supporting quantum programming have been proposed, such as \texttt{Qiskit}, \texttt{Cirq}, \texttt{Q\#}, and \texttt{ProjectQ}. We can use these frameworks to develop quantum software.
Here is an example of \texttt{Qiskit} programs\footnote{\url{https://github.com/Qiskit/qiskit-terra/blob/main/examples/python/ibmq/hello_quantum.py}}:

\begin{figure}[h]
\begin{CodeOut}
\footnotesize{
\begin{alltt}
1    \textbf{from} qiskit \textbf{import} QuantumCircuit
2    \textbf{from} qiskit \textbf{import} execute, IBMQ, BasicAer
3    qc = \textbf{QuantumCircuit}(2, 2)
4    qc.\textbf{h}(0)
5    qc.\textbf{cx}(0, 1)
6    qc.\textbf{measure}([0, 1], [0, 1])
7    sim = BasicAer.\textbf{get_backend}("qasm_simulator")
8    job_sim = \textbf{execute}(qc, sim)
9    result_sim = job_sim.\textbf{result}()
10   \textbf{print}(result_sim.\textbf{get_counts}(qc))
\end{alltt}
    }
    \end{CodeOut}
    \caption{A Qiskit Program.}
    \label{fig:A Qiskit Program.}
    \vspace*{-4mm}
\end{figure}

This program aims to create a Bell state. In lines 1-2, it imports related packages. Line 3 creates a quantum circuit acting on one quantum register of two quantum bits (qubits) and one classical register of two classical bits. Line 4 adds an H gate on qubit 0, putting this qubit in superposition. Line 5 adds a CX (CNOT) gate on control qubit 0 and target qubit 1, putting the qubits in a Bell state. Line 6 measures the qubits 0 and 1 of the quantum register, respectively, and puts the values into the corresponding classical bits 0 and 1 of the classical register. Line 8 executes the circuit on the QASM simulator.

\section{Methodology}
\label{sec:methodology}
\subsection{Dataset}
\begin{figure}[h]
    \centering
    \includegraphics[width=0.4\textwidth]{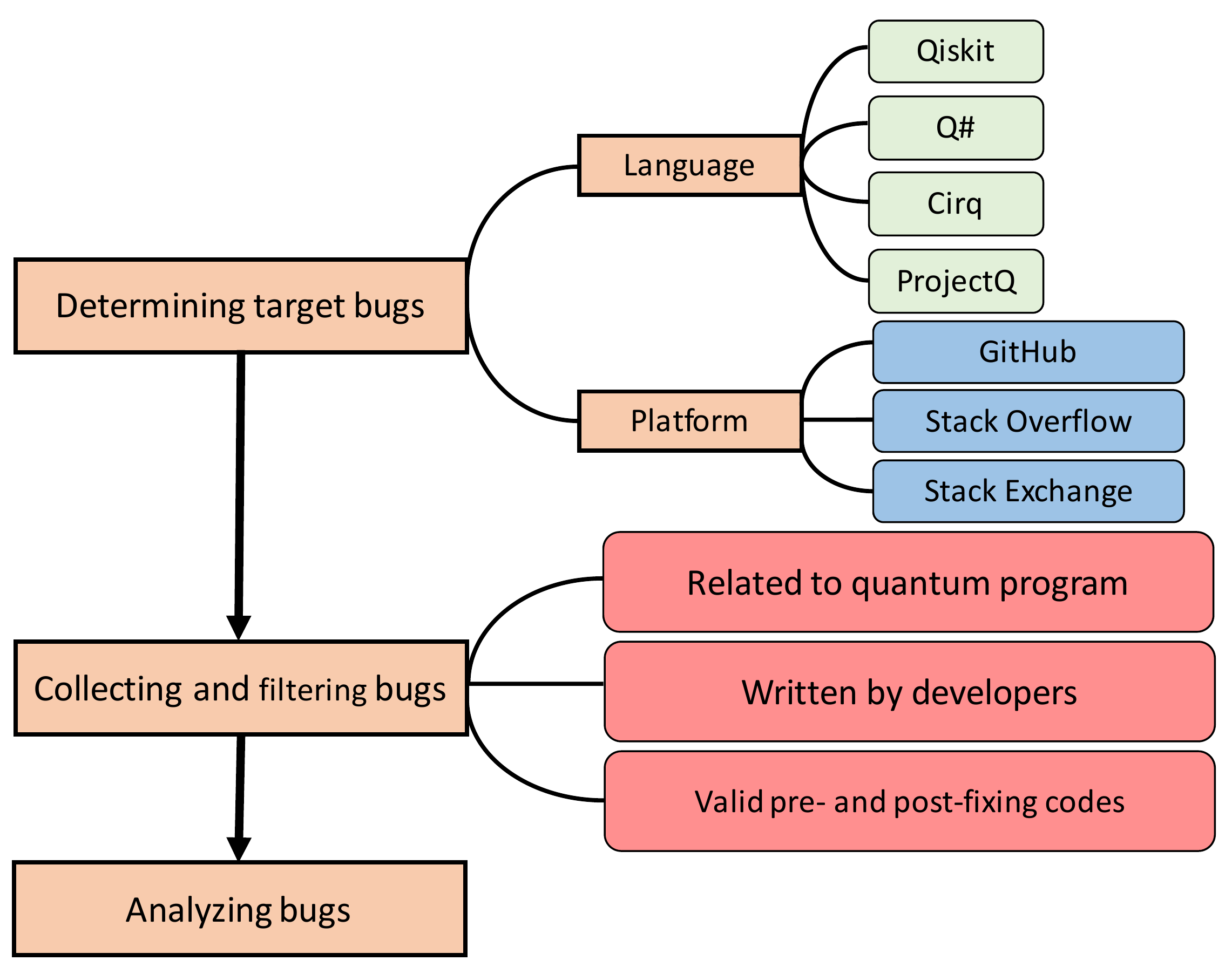}
    \caption{The process of collecting and analyzing bugs}
    \label{fig:process}
    \vspace*{-2mm}
\end{figure}

The data collection and analysis workflow is divided into the following steps: determining target bugs, collecting and filtering bugs, and analyzing bugs. Figure \ref{fig:process} depicts our data collection and analysis process.

\subsubsection{Determining target bugs}In order to collect as many bugs as possible, we collected bugs generated in the most commonly used quantum programming languages, including \texttt{Qiskit}, \texttt{Cirq}, \texttt{Q\#}, and \texttt{ProjectQ}. We collected bugs raised on GitHub, Stack Overflow, and Stack Exchange in terms of platforms for reporting bugs. 

\subsubsection{Collecting and filtering bugs} We manually collect bugs for each target language on different platforms since the number of bugs in quantum languages is not very large, and we need to filter the bugs to find those that meet our requirements. There are three filter conditions for bugs as follows:

\begin{itemize}[leftmargin=2em]
\item\textbf{Related to quantum programs. }We only collect bugs related to quantum programs. If a bug is in a program that does not involve any quantum computation or operation, then the bug will not be included in the collection. It is important to note that although we require the bug to occur in a quantum program, this does not mean the bug itself is quantum-related. Instead, it may be a bug that would also occur in classical programming languages, which we will specifically discuss later.

\item\textbf{Written by developers. }We will only work on analyzing bugs written by developers and will not discuss bugs that occur in the quantum programming language itself. This will have little impact on the bug collection on Stack Overflow and Stack Exchange since the issues on these platforms are raised by developers. However, on GitHub, only a few issues are related to developers. The vast majority of issues are specific to the quantum language itself.

\item\textbf{Valid pre- and post-fixing codes.} To facilitate analysis, we require that the collected bugs have pre- and post-fixing code. Also, we need to confirm that the changes are effective, depending on the feedback from the user who raised the bug.
\end{itemize}

\subsubsection{Analyzing bugs}We analyze all the valid data collected. We count the distribution of bugs for different research questions, focus on representative examples, and finally give our findings.

\begin{table}[t]
\caption{\label{table:TotalDataset} Repositories considered in the study with bugs in quantum programs}
\footnotesize{
\begin{center}
\renewcommand\arraystretch{1.0}
\begin{tabular}{lll}
\hline
Repository & Total & Fix \\ \hline
Qiskit     & 138   & 71  \\
Cirq       & 25    & 16  \\
Q\#        & 21    & 7   \\
ProjectQ   & 3     & 2\\ \hline
Total      & 187   & 96  \\ \hline
\end{tabular}
\end{center}
}
\vspace*{-4mm}
\end{table}

\subsection{Research Question}
Our study is driven by the following research questions: 

\subsubsection{RQ1} \textit{Does this bug only occur in quantum programs?}
Most quantum languages are not completely new. for example, the major versions of \texttt{Cirq} and \texttt{Qiskit} use the \texttt{Python} programming language, while \texttt{Q\#} is based on \texttt{C\#} and \texttt{F\#} extensions. In such cases, bugs in the fixes may come from classical programming languages or come specifically from quantum languages. Studying this issue helps us to better understand to what extent the repair of quantum programs can benefit from the existing knowledge of classical software repair. Our study investigates the proportion of classical bugs and quantum-specific bugs in quantum programs.

\subsubsection{RQ2} \textit{Where is the location of the bug in the programs?}
Similar to classical programs, to fix bugs in quantum programs, we first need to investigate the distribution of bugs. Zhong and Su \cite{7194637Realbug} show that bugs in a program can occur outside the source file, so knowing where the bugs are located can help us to find the focus of bug fixes. They also indicate that more than one line of code may need to be modified to fix a bug. The existence of dependencies in the fixed code affects the difficulty of fixing the code. Our study analyzes the distribution of bugs under real quantum programs. The results of the study help locate bugs in quantum programs.

\subsubsection{RQ3} \textit{How complex to fix bugs in quantum programs?}
In addition to analyzing the location of bugs, it is also essential to understand the complexity of bugs to fix bugs in quantum programs automatically. When there is more than one bug in a program, dependencies between bugs will impact the difficulty of fixing the bug. For bugs with dependencies, breaking the dependencies during the fixing process will lead to the failure of the bug fixing. In our study, we analyze the dependencies of bug codes in quantum programs following \cite{7194637Realbug}, and describe the complexity in four levels.

\subsubsection{RQ4} \textit{Are there bug patterns in the quantum program?}
We found that some bugs overlapped and repeated by reproducing them in the corresponding quantum programs. Intuitively, we infer that they can be distributed over some universal patterns or regularities, and this information would be helpful to guidance for fixing quantum-specific bugs. Based on this, we performed a thorough investigation of these bugs and classified them by several principles. 

\section{Empirical Results}
\label{sec:result}

\subsection{RQ1: Classical or Quantum-Specific Bugs}
This research question investigates the proportion of classical bugs and quantum-specific bugs in quantum software. Generally, quantum-specific bugs are different from classical software bugs as repairing them requires the domain knowledge of quantum-related concepts, properties, computational formulas, and quantum programming languages. Table \ref{table:ClassicalQuantum} shows the statistics of the number of classical and quantum-specific bugs in the data we collected. From the results of the investigation, we can derive the following findings: 

\begin{table}[h]
\centering
\caption{\label{table:ClassicalQuantum}Number of classical and quantum specific bugs}
\begin{tabular}{ll}
\hline
Type      & Count \\ \hline
Classical & 17    \\
Quantum   & 79    \\ \hline
Total     & 96    \\ \hline
\end{tabular}
\vspace*{-0mm}
\end{table}

\textit{\textbf{Finding 1.}} Clearly, the number of quantum-specific bugs is much greater than the number of classical bugs (more than 80\%). Since most of the existing bug fixing approaches focus on classical programs, it will be a great challenge to deal with the bugs in quantum programs. It also means that researching bug fixing in quantum programs is highly worthwhile.

\subsection{RQ2: Bug Distribution}
We investigated whether files other than the source files were modified during the fixing of the quantum program. The results of the investigation lead to the following findings: 

\textit{\textbf{Finding 2.}} In all the bugs we collected, only the source files have been modified. We believe there are two reasons for this situation: the first is that the difficulty of simulating large quantum programs by classical computers has led to the fact that quantum programs, in general, are limited to simpler functions and therefore do not need other files to constitute the project. Secondly, we can find that programmers wrote a significant part of the programs in our data collection exploring the quantum language. For example, some programs contain only attempts on how to add the \texttt{QFT} circuit to the code. This type of quantum program would have only source files. 

\begin{figure}[h]
    \flushleft
    \includegraphics[width=0.4\textwidth]{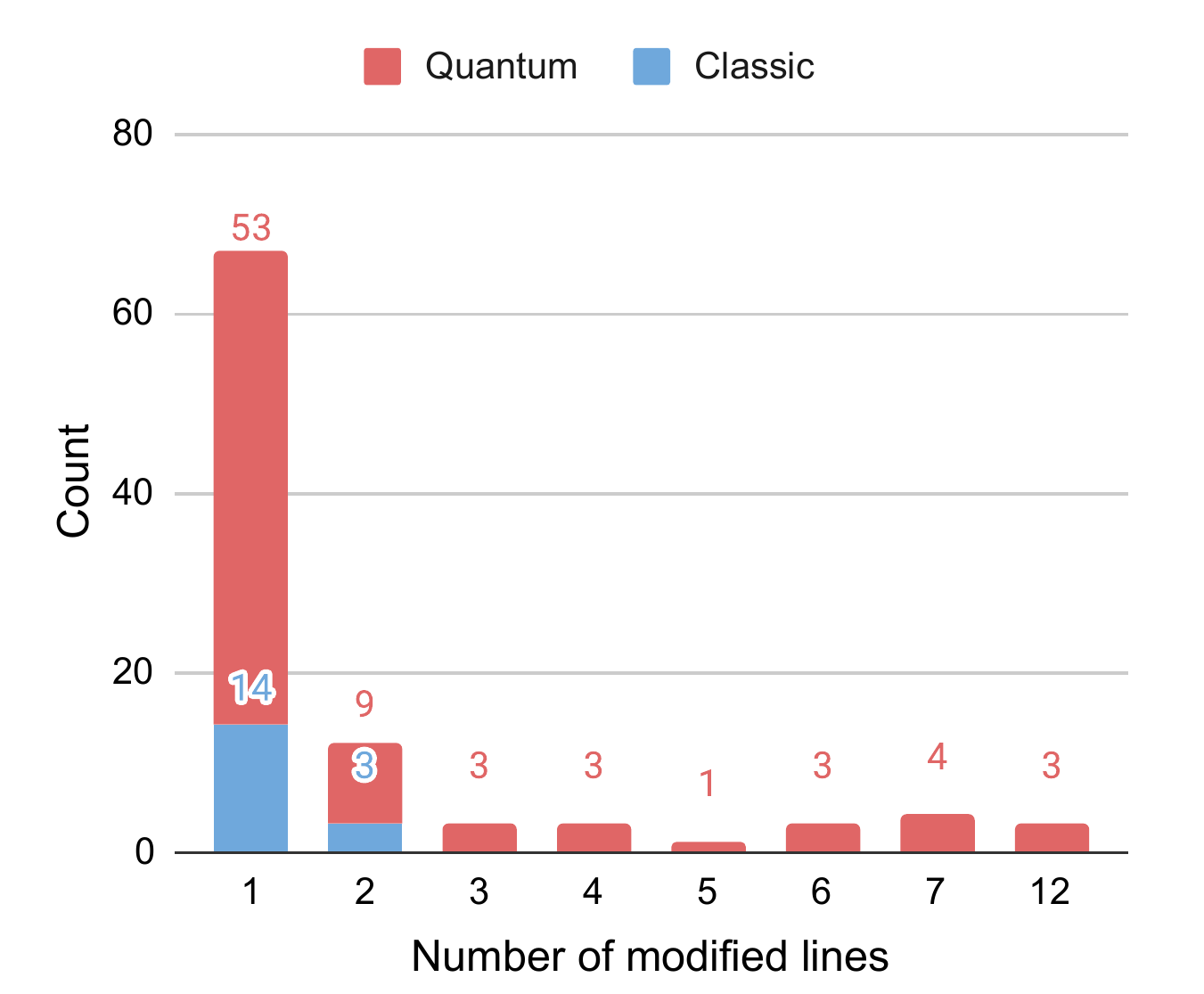}
    \caption{The distribution of the number of lines that need to be modified for quantum-specific bugs and classical bugs in our collected data.}
    \label{fig:modifiedline}
    \vspace*{-2mm}
\end{figure}

In addition, we counted the number of lines of code that needed to be modified to fix each bug in the quantum program. We define the action of adding, deleting, and updating a line of code as a single modification. Figure \ref{fig:modifiedline} shows the statistics of the number of lines needed to fix classical bugs and quantum-specific bugs in quantum programs, respectively. The stacked bars show the number of bugs, red for quantum-specific bugs while blue for classical bugs. Furthermore, the horizontal coordinates in the chart indicate the number of lines that need to be modified.

\textit{\textbf{Finding 3.}} Overall, more than 70\% of the bugs in our collection of quantum programs can be fixed by modifying only one line of code, while less than 10\% of the code requires modifying more than six lines. This implies that the cost of fixing bugs is generally low in existing quantum programs.

\textit{\textbf{Finding 4.}} From the type of bugs, we can find that the number of lines to be modified for classical bugs appearing in quantum programs is generally low (at most, two lines of code need to be modified). In Zhong and Su's work\cite{7194637Realbug}, less than 30\% of bugs in classical programs can be fixed by only one repair action. We can see that it is significantly less challenging to fix classical bugs in quantum programs than in classical programs. So more attention should be focused on how to fix quantum-specific bugs in quantum programs.

\subsection{RQ3: Bug Complexity} 
\begin{figure}[h]
    \flushleft
    \includegraphics[width=0.4\textwidth]{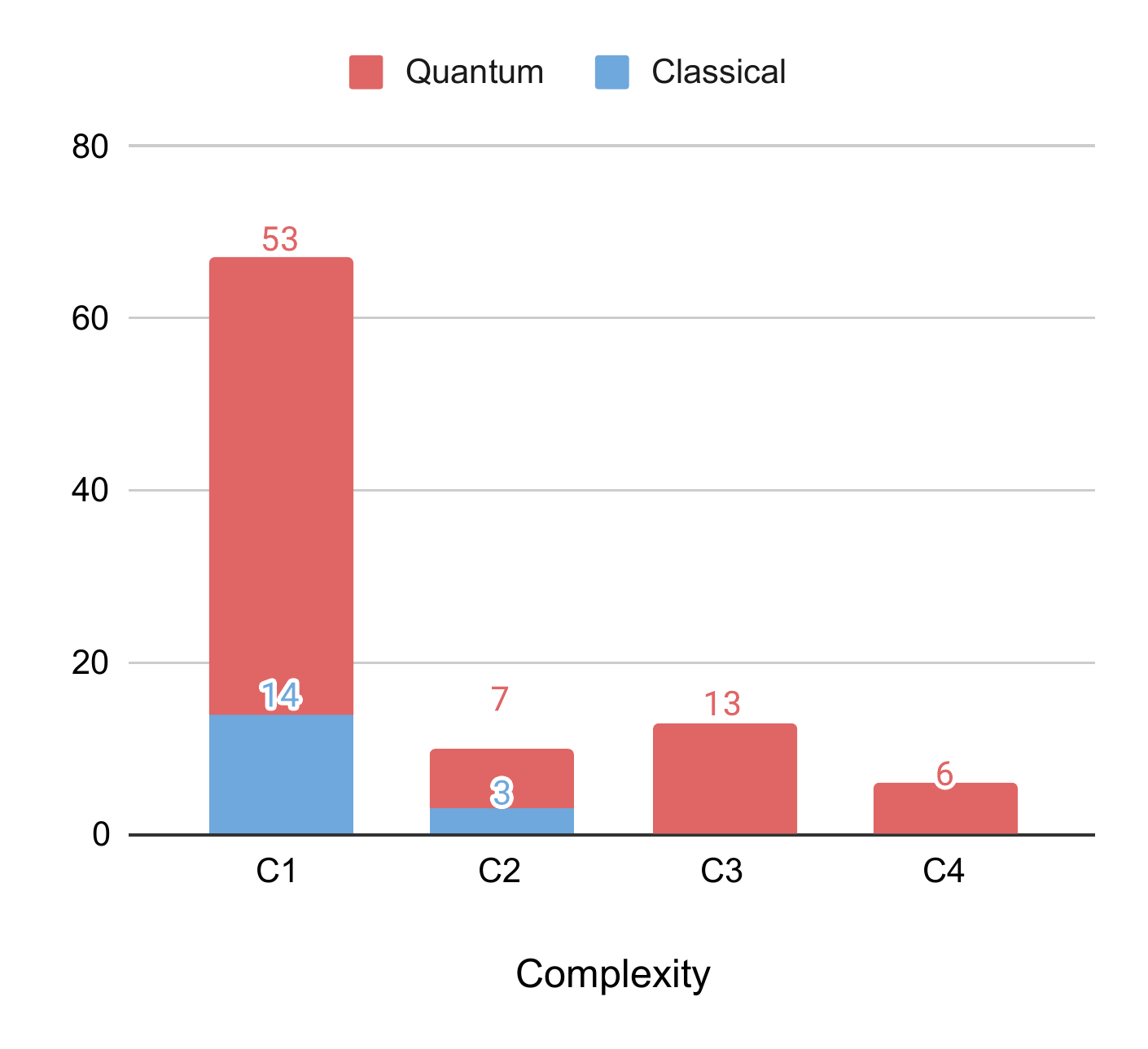}
    \caption{The distribution of complexity. Classic bugs and quantum-specific bugs are shown in the red and blue bar respectively. }
    \label{fig:Complexity}
    \vspace*{-4mm}
\end{figure}

To explore the difficulty of fixing bugs in quantum programs, we need a more detailed definition of the complexity of bugs in quantum programs. As one of the well-established methods, Paltenghi and Pradel \cite{paltenghi2021bugs} determine the complexity of bug fixing in terms of the number of lines of code that are in error. While such an approach can quickly determine the occurrence of bugs that some automatic fixes cannot handle (e.g., bugs that require fixing more than 20 lines), it is difficult to distinguish the difficulty of fixing bugs with fewer lines (e.g., whether there are dependencies between the data being fixed). Here, we refer to the four levels of fault complexity proposed by Zhong and Su \cite{7194637Realbug}: a single repair action (C1), non-data dependent repair actions (C2), data-dependent repair actions (C3), and mixture repair actions (C4). We consider the modification of a line as a repair action and thus apply this complexity classification to our work. Although this complexity classification was used in the paper for classical programs, we argue that this approach can also be applied to quantum programs in our study. Figure \ref{fig:Complexity} shows the results of our analysis, where we represent the classical bugs and quantum-specific bugs in each complexity with blue and red bars, respectively.

\textit{\textbf{Finding 5.}} Since single line fixes are directly classified into C1 complexity, the proportion of C1 complexity is much higher than the other three complexity levels in quantum programs. This reflects that the complexity of bugs in current quantum programs is generally low, but there are still some bugs that are difficult to fix (19 bugs of C3 and C4 complexity in total), and all the bugs in this part belong to quantum specific bugs, and thus the difficulty in bug fixing cannot be ignored.

\textit{\textbf{Finding 6.}} 
In our study, the cause of the errors in 5 of the 19 samples with complexity C3 and C4 is due to an incorrect implementation method. This error is usually because the programmer does not know the exact implementation of the required functionality, so more code manipulation is usually required to fix it. Moreover, this type of error is difficult for bug-fix automation approaches to work around because we cannot expect them to anticipate the functionality that the programmer needs to implement to make the correct fix. Furthermore, this means that in the development of existing quantum programs, there are fewer cases of C3 and C4 complexity errors if the programmer is known about the exact implementation of the desired function.

\subsection{RQ4: Bug Patterns}

Our study also analyzed and counted the bug patterns of the 96 valid samples collected. We refer to the bug pattern classification used in the work of Paltenghi and Pradel \cite{paltenghi2021bugs}. Although their work is specific to quantum computing platforms, we believe that some of the bug patterns can also be used to study quantum program bugs. In addition, we have defined some new bug patterns for quantum programs to complement and improve the shortcomings of their bug patterns when applied to quantum programs. Our study divided the bug pattern into four major categories: API-related, incorrect application logic, math-related, and others, and also detailed several subcategories under each major category. To adapt it to the analysis of bugs in quantum programs, we further subdivide incorrect numerical computation under the math-related category by adding bug patterns specific to quantum programs, such as an incorrect gate. Table \ref{table:Bugpattern} shows the details of each bug pattern classification in our study and lists the number of bug samples under each bug pattern. The following discussion will briefly describe each bug pattern and detail our proposed additional quantum program-specific bug patterns.

\begin{table}[h]
\centering
\caption{\label{table:Bugpattern} Bug patterns in the investigated data}
\small
\resizebox{.95\columnwidth}{!}{
\begin{tabular}{lll|l}
\hline
Bug pattern                                                                                  &                                                                                                                 &                                                                        & Count \\ \hline
\multicolumn{1}{l|}{API-related}                                                             & API misuse                                                                                                      &                                                                        & 30    \\ \cline{2-4} 
\multicolumn{1}{l|}{}                                                                        & Outdated API client                                                                                             &                                                                        & 5     \\ \hline
\multicolumn{1}{l|}{\begin{tabular}[c]{@{}l@{}}Incorrect\\ application\\ logic\end{tabular}} & \multicolumn{1}{l|}{\multirow{2}{*}{\begin{tabular}[c]{@{}l@{}}Intermediate \\ representation\end{tabular}}}    & Missing information                                                    & 4     \\ \cline{3-4} 
\multicolumn{1}{l|}{}                                                                        & \multicolumn{1}{l|}{}                                                                                           & Wrong information                                                      & 6     \\ \cline{2-4} 
\multicolumn{1}{l|}{}                                                                        & \multicolumn{1}{l|}{\multirow{2}{*}{\begin{tabular}[c]{@{}l@{}}Refer to wrong \\ program element\end{tabular}}} & Wrong concept                                                          & 2     \\ \cline{3-4} 
\multicolumn{1}{l|}{}                                                                        & \multicolumn{1}{l|}{}                                                                                           & Wrong identifier                                                       & 1     \\ \cline{2-4} 
\multicolumn{1}{l|}{}                                                                        & Incorrect scheduling                                                                                            &                                                                        & 1     \\ \cline{2-4} 
\multicolumn{1}{l|}{}                                                                        & Wrong declaration                                                                                               &                                                                        & 1     \\ \cline{2-4} 
\multicolumn{1}{l|}{}                                                                        & \multicolumn{2}{l|}{Wrong way of implementation}                                                                                                                                         & 7     \\ \cline{2-4} 
\multicolumn{1}{l|}{}                                                                        & \multicolumn{1}{l|}{Qubit-related}                                                                              & Incorrect qubit order                                                  & 4     \\ \cline{3-4} 
\multicolumn{1}{l|}{}                                                                        & \multicolumn{1}{l|}{}                                                                                           & Incorrect qubit count                                                  & 1     \\ \hline
\multicolumn{1}{l|}{Math-related}                                                            & \multicolumn{1}{l|}{\multirow{5}{*}{\begin{tabular}[c]{@{}l@{}}Incorrect numerical\\ computation\end{tabular}}} & Incorrect gate                                                         & 8     \\ \cline{3-4} 
\multicolumn{1}{l|}{}                                                                        & \multicolumn{1}{l|}{}                                                                                           & \begin{tabular}[c]{@{}l@{}}Incorrect matrix\\ computation\end{tabular} & 9     \\ \cline{3-4} 
\multicolumn{1}{l|}{}                                                                        & \multicolumn{1}{l|}{}                                                                                           & Incorrect initial state                                                & 2     \\ \cline{3-4} 
\multicolumn{1}{l|}{}                                                                        & \multicolumn{1}{l|}{}                                                                                           & Incorrect measurement                                                  & 3     \\ \cline{3-4} 
\multicolumn{1}{l|}{}                                                                        & \multicolumn{1}{l|}{}                                                                                           & \begin{tabular}[c]{@{}l@{}}Potential all-zeros\\ matrix\end{tabular}   & 1     \\ \cline{2-4} 
\multicolumn{1}{l|}{}                                                                        & \multicolumn{2}{l|}{Incorrect randomness handling}                                                                                                                                       & 3     \\ \hline
\multicolumn{1}{l|}{Others}                                                                  & \multicolumn{2}{l|}{Misconfiguration}                                                                                                                                                    & 4     \\ \cline{2-4} 
\multicolumn{1}{l|}{}                                                                        & \multicolumn{2}{l|}{Type problem}                                                                                                                                                        & 1     \\ \cline{2-4} 
\multicolumn{1}{l|}{}                                                                        & \multicolumn{2}{l|}{Typo}                                                                                                                                                                & 2     \\ \cline{2-4} 
\multicolumn{1}{l|}{}                                                                        & \multicolumn{2}{l|}{Overflow error}                                                                                                                                                      & 1     \\ \hline
\end{tabular}}
\end{table}

\subsubsection{API-related} In quantum programs, if the error is caused by an API, we classify it into \textit{API-related} bug pattern. There are two main subcategories under this category, including \textit{API misuse}, which refers to errors caused by using the wrong function method or introducing the wrong parameters into a function, and \textit{outdated API client}, which refers to code edited with an older, deprecated version of the API. Figure \ref{fig:APImisuse} shows a bug belonging to the API-misuse, where the \textbf{execute} function got an unexpected keyword argument \texttt{seed}.

\begin{figure*}[h]
\begin{center}
\begin{CodeOut}
\footnotesize{
\begin{alltt}
    shots = 1024
-   job = \textbf{execute}(circuits, backend = Aer.\textbf{get_backend}('qasm_simulator'), shots=shots, seed=8)
+   job = \textbf{execute}(circuits, backend=Aer.\textbf{get_backend}('qasm_simulator'), shots=shots)
    result = job.\textbf{result}()
\end{alltt}
    }
    \end{CodeOut}
    \caption{Example of the API misuse, in which the programmer used an unexpected keyword argument \texttt{seed}.}
    \label{fig:APImisuse}
\end{center}
\end{figure*}

\subsubsection{Incorrect application logic} This type of bug pattern is mainly caused by errors in the implementation of the application logic when writing a quantum program. This bug pattern consists of six subcategories: \textit{Intermediate representation} is a bug caused by the corruption or loss of an intermediate variable in a quantum program. \textit{Referring to the wrong program element} is an error caused by a programmer confusing two elements (e.g., two related concepts and two identifiers with similar names) in a quantum program and misusing them. \textit{Incorrect scheduling} is a bug caused by code running at the wrong timestamp in a quantum program (e.g., a new instruction is incorrectly scheduled to execute on the hardware when it is occupied). In \textit{qubit-related}, it mainly includes qubit-related errors, such as counting the wrong number of the qubits or causing wrong ordered qubits. Figure \ref{fig:wrongconcept} shows an example that the programmer confused the two operations in \texttt{Qiskit}: \texttt{Gate}s, which can be controlled, and \texttt{Instruction}s, which cannot be controlled, leading to the error.
Besides, we define two new bug patterns for existing quantum programs, which are discussed in detail as follows.

\begin{figure}[h]
\begin{CodeOut}
\footnotesize{
\begin{alltt}
    qc = \textbf{QuantumCircuit}(3)
    outer_level = \textbf{QuantumCircuit}(2, name='outer')
    inner_level = \textbf{QuantumCircuit}(2, name='inner')
    inner_level.\textbf{x}(0)
-   outer_level.\textbf{append}(inner_level, [0,1])
+   outer_level.\textbf{append}(inner_level.\textbf{to_gate}(), [0,1])
    qc.\textbf{append}(outer_level.\textbf{control}(), [0,1,2])
\end{alltt}
    }
    \end{CodeOut}
    \caption{Example of the wrong concept, in which the \texttt{inner\_level} produced from \textbf{QuantumCircuit} cannot be controlled. }
    \label{fig:wrongconcept}
\end{figure}

\textit{Wrong declaration.} This bug pattern can happen mainly in \texttt{Q\#}. Since the code is automatically wrapped in the namespace in \texttt{Q\#} \texttt{Jupyter notebooks}, there is no need to re-declare the same namespace in the code. To fix this bug, one needs to remove the extra namespace declaration from the code. 

\textit{Wrong way of implementation.} In the process of developing quantum programs, this type of bug can be caused if the programmer is not clear about the exact implementation of the function needed. This bug pattern is characterized by a large amount of code to be fixed and its high complexity (5 out of 7 samples need to be fixed with more than 5 lines, with a complexity of C3 or C4), which is a great challenge for the current bug fix automation approaches.

\subsubsection{Math-related} In developing a quantum program, the programmer usually has to manipulate the qubits through a series of mathematical operations to accomplish the desired function. In Paltenghi and Pradel's work\cite{paltenghi2021bugs}, math-related bugs are divided into two categories: \textit{incorrect numerical computations} and \textit{incorrect randomness handling}. If the code misuses randomness or probability, it will be classified into incorrect randomness handling patterns. Incorrect numerical computations mean that the formula is misused. In quantum programs, since the operations on qubits are more diverse than those in classical programs, in our study, we further subdivide this bug pattern as follows.

\textit{Incorrect gate. }In quantum programs, to control quantum circuits, programmers usually need to use quantum logic gates to process qubits to implement a particular quantum algorithm. In this process, if the wrong quantum logic gate is used, or if the wrong qubits are processed with the quantum logic gate, it can lead to anomalies in the quantum program. Figure \ref{fig:incorrectgate} gives an example of an incorrect gate, in which the output does not match the expected result because the programmer used the wrong quantum logic gate.

\begin{figure}[h]
\begin{CodeOut}
\footnotesize{
\begin{alltt}
    circuit = \textbf{QuantumCircuit}(2)
    circuit.\textbf{h}(0)
-   circuit.\textbf{h}(1)
+   circuit.\textbf{x}(1)
    circuit.\textbf{cx}(0,1)
    circuit.\textbf{measure_all}()
\end{alltt}
    }
    \end{CodeOut}
    \caption{Example of the incorrect gate usage resulting in an output that is not as expected.}
    \label{fig:incorrectgate}
\end{figure}

\textit{Incorrect initial state. }In quantum programs, when writing quantum algorithms, the programmer needs to set the initial state of the qubits. If the wrong initial state is used, it can lead to an error in the quantum circuit. In Figure \ref{fig:incorrectinitial}, the programmer tries to create the Bell State $\frac{1}{\sqrt{2}}\left(\vert01\rangle + \vert10\rangle\right)$, however, the wrong initial state is used, leading to the wrong result $\frac{1}{\sqrt{2}}\left(\vert11\rangle - \vert00\rangle\right)$.

\begin{figure}[h]
\begin{CodeOut}
\footnotesize{
\begin{alltt}
-   sv = Statevector.\textbf{from_label}('01')
+   sv = Statevector.\textbf{from_label}('10')
    mycircuit = \textbf{QuantumCircuit}(2)
    mycircuit.\textbf{h}(0)
    mycircuit.\textbf{cx}(0,1)
\end{alltt}
    }
    \end{CodeOut}
    \caption{Example of the incorrect initial state resulting in a wrong expression of the Bell State.}
    \label{fig:incorrectinitial}
\end{figure}

\vspace*{2mm}
\textit{Incorrect measurement. }In quantum programs, measurement is a common operation in quantum algorithms, but it can cause errors in quantum programs if measurements are made at the wrong time or on the wrong qubits. In Figure \ref{fig:incorrectmeasurement}, the programmer reversed the order of the qubits when measuring them, leading to an error in the measurement result. 

\begin{figure}[h]
\begin{CodeOut}
\footnotesize{
\begin{alltt}
-   circuit.\textbf{measure}([0,1,2], [0,1,2])
+   circuit.\textbf{measure}([0,1,2], [2,1,0])
\end{alltt}
    }
    \end{CodeOut}
    \caption{Example of wrong measurement order leading to incorrect results.}
    \label{fig:incorrectmeasurement}
\end{figure}

\textit{Potential all-zeros matrix. }In general, matrix operations frequently occur in quantum programs. And in matrix operations, unintentionally generated all-zeros matrices pose a risk to the operation and cause the algorithm's stability to deteriorate. Figure \ref{fig:potentialallzeros} shows a bug belonging to this bug pattern, where the operation \texttt{.\textbf{dot}(unitary[randnum[0]] - id)} causes the case where \texttt{rho} is an all-zeros matrix occurs frequently.

\begin{figure*}[h]
\begin{center}
\begin{CodeOut}
\footnotesize{
\begin{alltt}
    state0 = np.\textbf{array}([[1,0],[0,0]])
    state1 = np.\textbf{array}([[0,0],[0,1]])
+   states = [state0, state1]
    record_rho = np.\textbf{zeros}([4,4])
    for i in range(snapshot_num):
        randnum = np.\textbf{random}.\textbf{randint}(0,3,size=2)
        result = \textbf{one_shot}(randnum)
+       bit0, bit1 = [\textbf{int}(x) for x in \textbf{list}(result.\textbf{keys}())[0]]
+       U0, U1 = unitary[randnum[0]], unitary[randnum[1]]
-       if result.\textbf{get}('00') == 1:
-           rho = np.\textbf{kron}(3*np.\textbf{dot}(unitary[randnum[0]].\textbf{conj}().T,state0).\textbf{dot}(unitary[randnum[0]] - id), 
                  3*np.\textbf{dot}(unitary[randnum[1]].\textbf{conj}().T,state0).\textbf{dot}(unitary[randnum[1]]) - id)
-       elif result.get('01') == 1:
-           rho = np.\textbf{kron}(3*np.\textbf{dot}(unitary[randnum[0]].\textbf{conj}.T,state0).\textbf{dot}(unitary[randnum[0]] - id),
                  3*np.\textbf{dot}(unitary[randnum[1]].conj().T,state1).\textbf{dot}(unitary[randnum[1]]) - id)
-       elif result.get('10') == 1:
-           rho = np.\textbf{kron}(3*np.\textbf{dot}(unitary[randnum[0]].\textbf{conj}.T,state1).\textbf{dot}(unitary[randnum[0]] - id),
                  3*np.\textbf{dot}(unitary[randnum[1]].\textbf{conj}.T,state0).\textbf{dot}(unitary[randnum[1]]) - id)
-       else:
-           rho = np.\textbf{kron}(3*np.\textbf{dot}(unitary[randnum[0]].\textbf{conj}.T,state1).\textbf{dot}(unitary[randnum[0]] - id),
                  3*np.\textbf{dot}(unitary[randnum[1]].\textbf{conj}.T,state1).\textbf{dot}(unitary[randnum[1]]) - id)
+       rho = np.\textbf{kron}(3* U0.\textbf{conj}.T @ states[bit0] @ U0 - id, 3*U1.\textbf{conj}.T @ states[bit1] @ U1 - id)
        record_rho = record_rho + rho
\end{alltt}
    }
    \end{CodeOut}
    \caption{Example of a bug that may cause all-zeros matrix because of the operation \texttt{.\textbf{dot}(unitary[randnum[0]] - id)}. }
    \label{fig:potentialallzeros}
\end{center}
\end{figure*}

\textit{Incorrect matrix computation. }Another common type of operation in quantum programs is matrix operations. Although the processing of qubits using quantum logic gates can also be considered a matrix operation, in our study, we distinguish between the two because, usually, quantum logic gates are pre-defined functions in quantum languages. In contrast, other matrix operations need to be implemented by the programmer in his code, and it is difficult to classify this part of the operation as some classical quantum logic gate. Figure \ref{fig:incorrectmatrix} gives an example of incorrect matrix computation, in which the simulator throws an error because the \texttt{U\_error} matrix set by the programmer is not unitary.

\begin{figure*}[h]
\begin{center}
\begin{CodeOut}
\footnotesize{
\begin{alltt}
+   z = 0.995004165 + 1j * 0.099833417
+   z = z / \textbf{abs}(z)
+   u_error = np.\textbf{array}([[1, 0], [0, z]])
    noise\_params = \{'U':
                         \{'gate_time': 1,
                        'p_depol': 0.001,
                        'p_pauli': [0, 0, 0.01],
-                       'U_error': [[[1, 0], [0, 0]], [[0, 0], [0.995004165, 0.099833417]]]\}
+                       'U_error': u_error \}
                    \}
\end{alltt}
    }
    \end{CodeOut}
    \caption{Example of an incorrect parameter of a matrix, where the matrix is non-unitary causing an error.}
    \label{fig:incorrectmatrix}
\end{center}
\end{figure*}

\subsubsection{Other bug patterns} In addition to the three main types of bug patterns mentioned above, there are some other patterns included in our study, which include four types of \textit{Misconfiguration}, \textit{Type problem}, \textit{Typo}, and \textit{Overflow error}. Figure \ref{fig:typeproblem} gives an example of type problem, in which the programmer misunderstands the return type of function \texttt{.\textbf{draw}(output = 'mpl')}. Since they are roughly indistinguishable from those in classical programs, we will not describe them in detail.

\begin{figure}[h]
\begin{CodeOut}
\footnotesize{
\begin{alltt}
    qr = \textbf{QuantumRegister}(2)
    cr = \textbf{ClassicalRegister}(2)
    Qc = \textbf{QuantumCircuit}(qr,cr)
    \textbf{print}('This is the initial state')
-   \textbf{print}(Qc.\textbf{draw}(output = 'mpl'))
+   Qc.\textbf{draw}(output = 'mpl')
\end{alltt}
    }
    \end{CodeOut}
    \caption{Example of the type problem caused by the misunderstanding of the return type of \texttt{.\textbf{draw}(output = 'mpl')}.}
    \label{fig:typeproblem}
\end{figure}

Based on the above analysis of bug patterns, we have the following findings: 

\textit{\textbf{Finding 7.}} Overall, the statistics show that in the existing quantum programs, programmers' bugs are mainly focused on \textit{API-related} bugs (35 out of 96 samples) and \textit{math-related} bugs (26 out of 96 samples). This means that programmers are still not proficient in quantum programming languages when developing quantum programs and have difficulties with the algorithms and manipulation of qubits. Therefore, it should be one of the focuses for future work on bug fixing.

\begin{figure*}[h]
    \centering
    \includegraphics[width=0.9\textwidth]{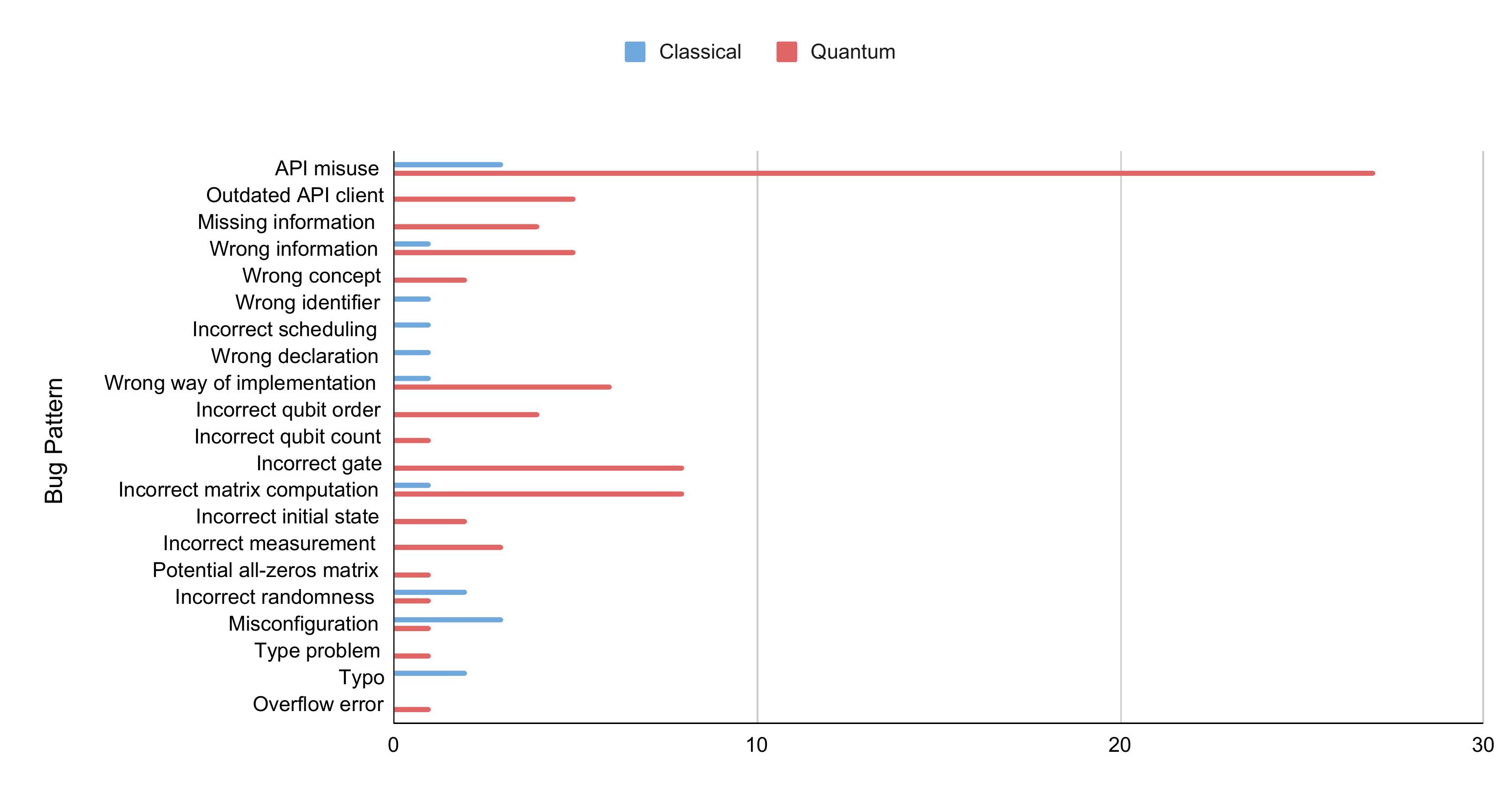}
    \caption{Bug pattern statistics for classical and quantum-specific bugs.}
    \label{fig:bugpattern}
    \vspace*{-4mm}
\end{figure*}

Our study also counts the cases of classical and quantum-specific bugs in quantum programs separately. The detailed data are shown in Figure \ref{fig:bugpattern}, where the vertical axis lists the bug patterns as mentioned above, and the horizontal axis shows the frequency of occurrence in the data we investigated, with the classical and quantum-specific bugs shown in blue and red bars, respectively.

\textit{\textbf{Finding 8.}} Consistent with the above findings, quantum-specific bugs are mainly concentrated in API-related and math-related bug patterns. On the other hand, classical bugs are smaller in total and evenly distributed across bug patterns, which means that in the study of bug fixing in quantum programs, researchers still need to pay extensive attention to the bugs that occur in classical programs. The high percentage of quantum-specific bugs in API-related and math-related bug patterns also indicates that the current bug fixing efforts are difficult to apply directly to quantum programs, and further research is needed in this area.

\section{Related Work}
\label{sec:related-work}
To the best of our knowledge, this work is the first study on the bug fixing of quantum programs. 
Many empirical studies on bug fixing have been proposed in classical software systems. Zhong and Su~\cite{7194637Realbug} collected more than 9000 real-world bugs from Java projects, analyzed them comparatively, and presented findings and insights. Zhang and Khomh {\it et al.}~\cite{zhang2012empirical} focused on the factors affecting bug fixing time. Park and Kim {\it et al.}~\cite{park2012empirical} discussed additional bug fixes, i.e., multiple attempts to fix bugs. Yin and Yuan {\it et al.}~\cite{yin2011fixes} researched bug fixes from large operating system code bases and proposed that the process of fixing bugs can also introduce bugs.
However, all the studies above focus on fixing the bugs of classical software systems rather than quantum software systems.

Paltenghi and Pradel~\cite{paltenghi2021bugs} presented an empirical study of bugs in quantum computing platforms. They summarized 223 real-world bugs in some open-source quantum computing platforms, conducted a detailed statistical analysis of bugs, and proposed some bug patterns specific to the field of quantum computing platforms. The difference between their work and ours is that they focused on the bugs on quantum computing platforms, while we focused on quantum programs themselves. However, the findings distilled from both studies can benefit the deep understanding of bugs related to quantum computing systems.  

In addition, several researches have been carried out  to study the bug patterns (types)~\cite{zhao2021identifying,huang2019statistical}, as well as bug benchmarks~\cite{campos2021qbugs,zhao2021bugs4q} in quantum programs, in order to support the debugging and testing of quantum software.

\section{Concluding Remarks}
\label{sec:conclusion}

In this paper, we have presented a comprehensive study of bug fixing in quantum programs. Based on the 96 bugs we collected in four popular quantum programming languages (\texttt{Qiskit}, \texttt{Cirq}, \texttt{Q\#}, and \texttt{ProjectQ}), we analyzed them in terms of the bug types, complexity, the file types being fixed, and their bug patterns, respectively, and distilled eight findings. Our research shows that a high percentage of bugs are quantum-specific bugs. Also, lack of proficiency in using quantum programming languages and miswritten quantum algorithms are key causal factors for bugs. We believe this study will help guide future research on automatically preventing and fixing bugs in quantum programs.

\bibliographystyle{IEEEtran}
\bibliography{IEEEabrv,qse-bibliography}

\end{document}